\documentclass{article}
\usepackage{graphicx} 
\usepackage{authblk}
\usepackage{amsmath}
\usepackage{geometry}
\usepackage{comment}
\usepackage[title]{appendix}
\usepackage[font={small}]{caption}
\usepackage{cite} 

\newcommand{\x}{\mathbf{x}}
\newcommand{\y}{\mathbf{y}}

\title{Improved cross-validated distances for multivariate pattern analysis}
\author[1,2]{Laurent Caplette}
\author[1,2]{Sarah Lippé}

\affil[1]{Department of Psychology, Université de Montréal}
\affil[2]{Azrieli Research Center of the CHU Sainte-Justine}
\date{\today}

\begin{document}

\maketitle

\begin{abstract}
Characterizing the dissimilarity of neural representations between experimental conditions, and tracking it across time, is a central goal of multivariate pattern analysis. Guggenmos et al. (2018) assessed the reliability of many of the measures that can be used for that purpose on MEG data and recommended the use of either the cross-validated Euclidean distance or the within-class-corrected Pearson distance. In this commentary, we show that we can improve upon these distances. First, we show that the cross-validated Euclidean distance is equivalent to a sum of between-partition distances and that this equivalence can be leveraged to obtain a generalized variant, with increased reliability and accuracy. Second, we use the relationship between Euclidean distance and Pearson correlation to define a cross-validated correlation distance in a similar way. The resulting distance is more accurate and interpretable than a formulation proposed by Guggenmos and colleagues. Finally, we discuss the relationship between our generalized cross-validation and within-class correction, another strategy often used to increase reliability, and we show that generalized cross-validation results in higher accuracy for the correlation distance.
\end{abstract}

\section{Introduction}

Multivariate pattern analysis (MVPA) is a popular method for characterizing the discriminability of brain activity patterns across different experimental conditions and uncovering which information is encoded in different brain regions. It has been used extensively to study the neural mechanisms underlying various perceptual and cognitive processes \cite{Haxby2001,Haxby2014,Kamitani2005,Kriegeskorte2006}. One widely used approach for MVPA is representational similarity analysis (RSA), in which distances (most commonly Euclidean or correlation distances) between activity patterns of different conditions are estimated \cite{Kriegeskorte2008,Kriegeskorte2013,Caplette2024,Cichy2014,Cichy2017}. Matrices of pairwise distances can then be related to ones obtained from computational models or other data modalities. Applying RSA to MEG data is particularly useful, as it makes it possible to investigate the neural dynamics underlying cognitive functions with high temporal resolution.

One important thing to keep in mind however, is that distances are susceptible to a noise bias, which can lead to less reliable estimates. Indeed, distances are non-negative and will typically increase with increasing noise, even when the true distance remains small. A potential strategy to mitigate this bias is cross-validation, in which brain activity pattern estimates from independent data partitions are used together to compensate for the noise. This strategy has been successful for the Euclidean distance: unlike its non-cross-validated counterpart, the cross-validated Euclidean distance is unbiased by noise. This enables performing additional analyses, including testing whether a measured distance is significantly different from 0. The cross-validated Euclidean distance (and the cross-validated Mahalanobis distance, which is its equivalent after multivariate noise normalization) has been used in multiple studies, to answer various scientific questions (e.g., \cite{Charest2018,Ejaz2015,Flounders2019,Guggenmos2018,Poncet2025,Walther2016}). Despite its success, we believe that, in most cases, small improvements can be made to the cross-validation equation. We detail these adjustments and how they improve reliability and accuracy in section 2. For the Pearson correlation distance, cross-validation cannot be used in the same way to eliminate noise bias but, if implemented correctly, it can be useful in other ways. In section 3, we build upon our previous adjustments and we define a new cross-validated Pearson correlation distance. We show that this new implementation eliminates the interpretability problem of the non-cross-validated correlation distance, where distances unexpectedly decrease after stimulus onset, rather than increase to reflect the greater discriminability between conditions. Moreover, compared to an alternative formulation proposed by Guggenmos et al. \cite{Guggenmos2018}, our implementation does not require regularization and is more accurate. Similarly to cross-validation, within-class correction, whereby distances computed between trials of the same condition are subtracted from between-condition distances, is also used to reduce noise bias. In section 4, we show that this strategy is in fact mathematically related to our method of cross-validation, but that our method leads to more accurate estimates for the Pearson correlation distance.

\section{A generalization of the cross-validated Euclidean distance}

The squared Euclidean distance can be defined in the following way:
\begin{equation}
d_{Euclidean}^2(\x,\y) = (\x-\y)^\top(\x-\y).
\end{equation}
Here, the vectors \(x\) and \(y\) can denote the brain activity patterns associated with two experimental conditions.

A ``cross-validated'' variant of that distance (henceforth simply referred to as the cross-validated, or C.V., Euclidean distance) can be derived by projecting the pattern vectors of one data partition onto those of an independent data partition \cite{Walther2016}:
\begin{equation}
d_{Euclidean,CV}^2(\x,\y) = (\x_A-\y_A)^\top(\x_B-\y_B),
\end{equation}
where \(A\) and \(B\) denote the two data partitions. 

We can show that such cross-validation is actually equivalent to the mean of between-condition between-partition distances minus the mean of within-condition between-partition distances (full derivation in Appendix A):
\begin{equation}
d_{Euclidean,CV}^2(\x,\y) = \frac{d^2(\x_A,\y_B) + d^2(\x_B,\y_A)}{2} - \frac{d^2(\x_A,\x_B) + d^2(\y_A,\y_B)}{2}.
\end{equation}

Using random partitions (which is usually feasible in event-related designs), between-condition between-partition distances are equivalent to between-condition within-partition distances. Indeed, partitions are arbitrary here, and the noise from one condition in one partition is as orthogonal to the noise of a different condition in another partition as it is to the noise of a different condition in the same partition. Therefore, within-partition distances can be included in the equation and the previous equation can be further generalized, while remaining completely unbiased:
\begin{equation}
d_{Euclidean,GCV}^2(\x,\y) = \frac{d^2(\x_A,\y_B) + d^2(\x_B,\y_A) + d^2(\x_A,\y_A) + d^2(\x_B,\y_B)}{4} - \frac{d^2(\x_A,\x_B) + d^2(\y_A,\y_B)}{2}.
\end{equation}
To clearly differentiate it from previous implementations of the cross-validated Euclidean distance, we define this distance as the Euclidean distance with generalized cross-validation (G.C.V.). Because it is defined from more individual distances, this formulation has the benefit of resulting in an estimator with reduced variance. Note also that this form of cross-validation is related to a different strategy of bias reduction: within-class correction (see section 4).

We applied this distance on the same data from Cichy et al. \cite{Cichy2014} that was used by Guggenmos and colleagues \cite{Guggenmos2018}. We found that our G.C.V. Euclidean distance was slightly more reliable than the existing C.V. Euclidean distance, that is, distances between conditions were more similar across different recording sessions. When testing its accuracy on simulated data with known ground truth, we found that the G.C.V. distance was also more accurate (i.e. closer to the ground truth distance) than its standard C.V. counterpart (Figure 1; see Appendix C for estimation and simulation details).

\begin{figure}
    \centering
    \includegraphics[width=0.8\linewidth]{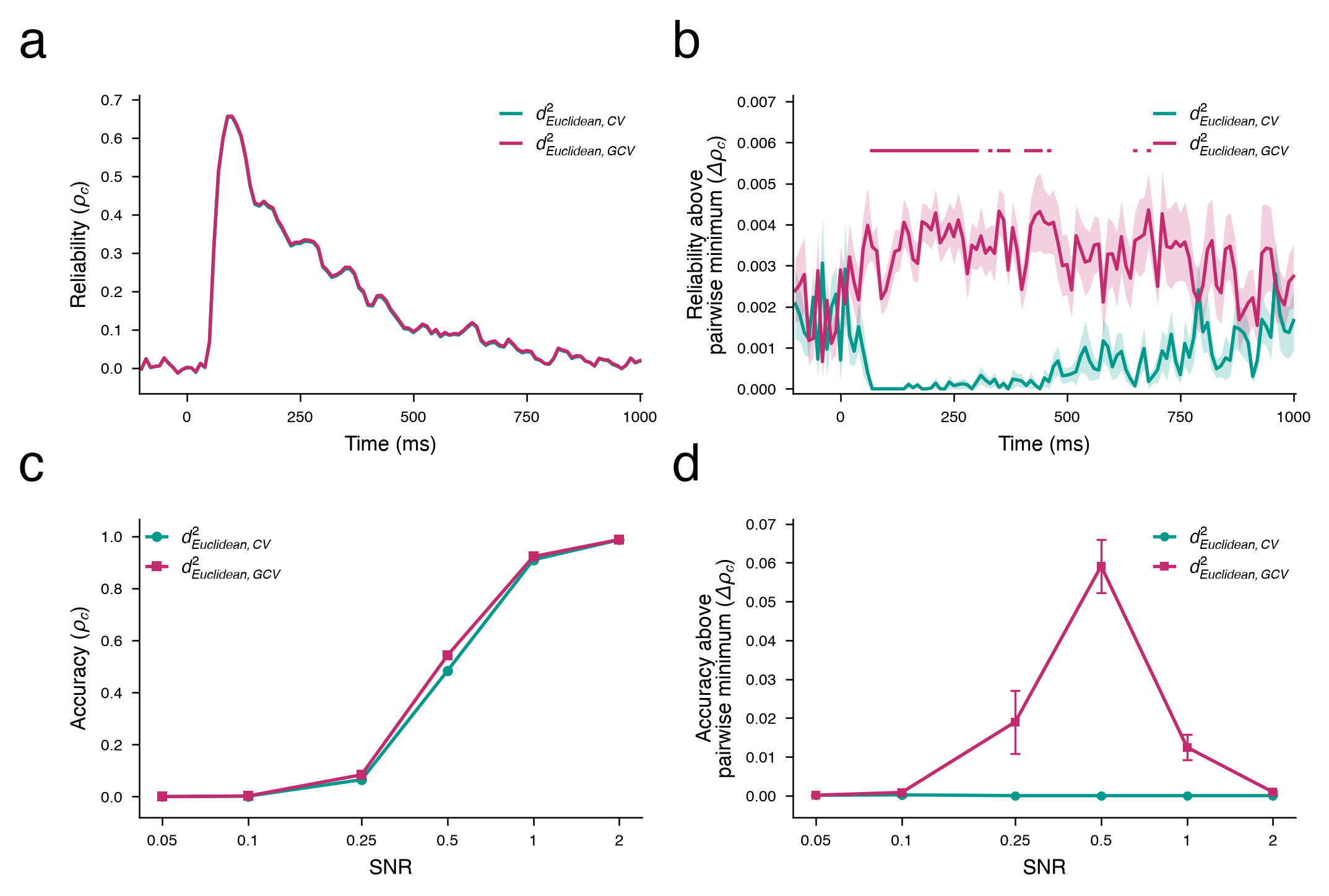}
    \caption{
    Reliability and accuracy of cross-validated Euclidean distances. \textbf{a)} Time-resolved distance reliability (MEG data). \textbf{b)} Between-distances reliability differences (MEG data). For each subject and time point, the minimum between the reliability values of both distances is taken and the difference to that minimum is computed. The group average is shown; the error bars illustrate the standard error of the mean and the top horizontal line illustrates significance ($p < .05$, FWER-corrected, sign permutation test). \textbf{c)} Distance accuracy across different SNRs (simulated data). \textbf{d)} Accuracy differences between distances (simulated data). For each SNR and simulation iteration, the minimum between the accuracy values of both distances is taken and the difference to that minimum is computed. The group average is shown and the error bars illustrate the standard deviation. See Appendix C for methodological details.
    }
    \label{fig:Fig1}
\end{figure}

\section{An improved cross-validated correlation distance}

The Pearson correlation distance is useful to get scale-insensitive estimates of dissimilarity. It can be defined as follows:
\begin{equation}
d_{Pearson}(\x,\y) = 1 - \frac{cov(\x,\y)}{\sqrt{var(\x)var(\y)}}.
\end{equation}
 
Guggenmos et al. \cite{Guggenmos2018} defined a cross-validated version of this distance by computing the covariance with vectors from two distinct data partitions, and generalizing the variances in the denominator to covariances between different data partitions:
\begin{equation}
d_{Pearson,CV,Guggenmos}(\x,\y) = 1 - \frac{\frac{1}{2}(cov(\x_A,\y_B)+cov(\x_B,\y_A))}{\sqrt{cov(\x_A,\x_B)cov(\y_A,\y_B)}}.
\end{equation}
As they note, this definition creates a practical problem: the covariances in the denominator can be very small or negative, leading to extremely positive, extremely negative, or even imaginary distances. Therefore, regularization must be applied to enforce a positive denominator. Guggenmos and colleagues enforced lower bounds for the covariances in the denominator and for the overall denominator, in addition to forcing the distance to lie between 0 and 2 (i.e. the bounds of the non-cross-validated correlation distance).

We argue that there is another way to define a cross-validated correlation distance that has more desirable properties. Indeed, the Pearson correlation distance is equivalent to a squared Euclidean distance between Z-transformed vectors, up to a scaling factor \cite{Berthold2016}. Therefore, we can cross-validate it in a similar way as is done for the squared Euclidean distance \cite{Walther2016} (full derivation in Appendix B):
\begin{equation}
d_{Pearson,CV,Ours}(\x,\y) = \frac{(Z(\x_A)-Z(\y_A))^\top(Z(\x_B)-Z(\y_B))}{2n},
\end{equation}
where \(Z(x)\) and \(Z(y)\) are centered and scaled versions of the original vectors, and \(n\) is the number of elements in each vector (the number of sensors in the case of MEG patterns).

Similarly to Eq.(3), this equation can be transformed to a sum of within-condition and between-condition correlations:
\begin{equation}
d_{Pearson,CV,Ours}(\x,\y) = \frac{r(\x_A,\x_B) + r(\y_A,\y_B)}{2} - \frac{r(\x_A,\y_B) + r(\x_B,\y_A)}{2},
\end{equation}
where \(r\) is the Pearson correlation coefficient.

Similarly to Eq.(4), assuming that partitions are random, we can further generalize:
\begin{equation}
d_{Pearson,GCV}(\x,\y) = \frac{r(\x_A,\x_B) + r(\y_A,\y_B)}{2} - \frac{r(\x_A,\y_B) + r(\x_B,\y_A) + r(\x_A,\y_A) + r(\x_B,\y_B)}{4}.
\end{equation}

Because there is no arbitrary regularization, this implementation is more accurate than the one described by Guggenmos et al. \cite{Guggenmos2018} (Figure 2e-f). Although their distance in its base form is slightly more reliable (significant difference at one time point; Figure 2a-b), this increased reliability is largely driven by the enforced regularization. At latencies where conditions are highly discriminable, the unregularized distance is more likely to go beyond the value of 2 and thus to be clipped at 2. Such clipping results in identical values across sessions, artificially increasing the distance's reliability. When this clipping manipulation is removed (Figure 2c-d), reliability decreases sharply, and we can see in fact that the G.C.V. correlation distance is more reliable most of the time. Another benefit of the G.C.V. correlation distance is that it has a more interpretable time course, with a mean distance starting around 0 and increasing after stimulus presentation, similarly to the within-class corrected distance (Figure 2g-h; see section 4).

\begin{figure}
    \centering
    \includegraphics[width=0.8\linewidth]{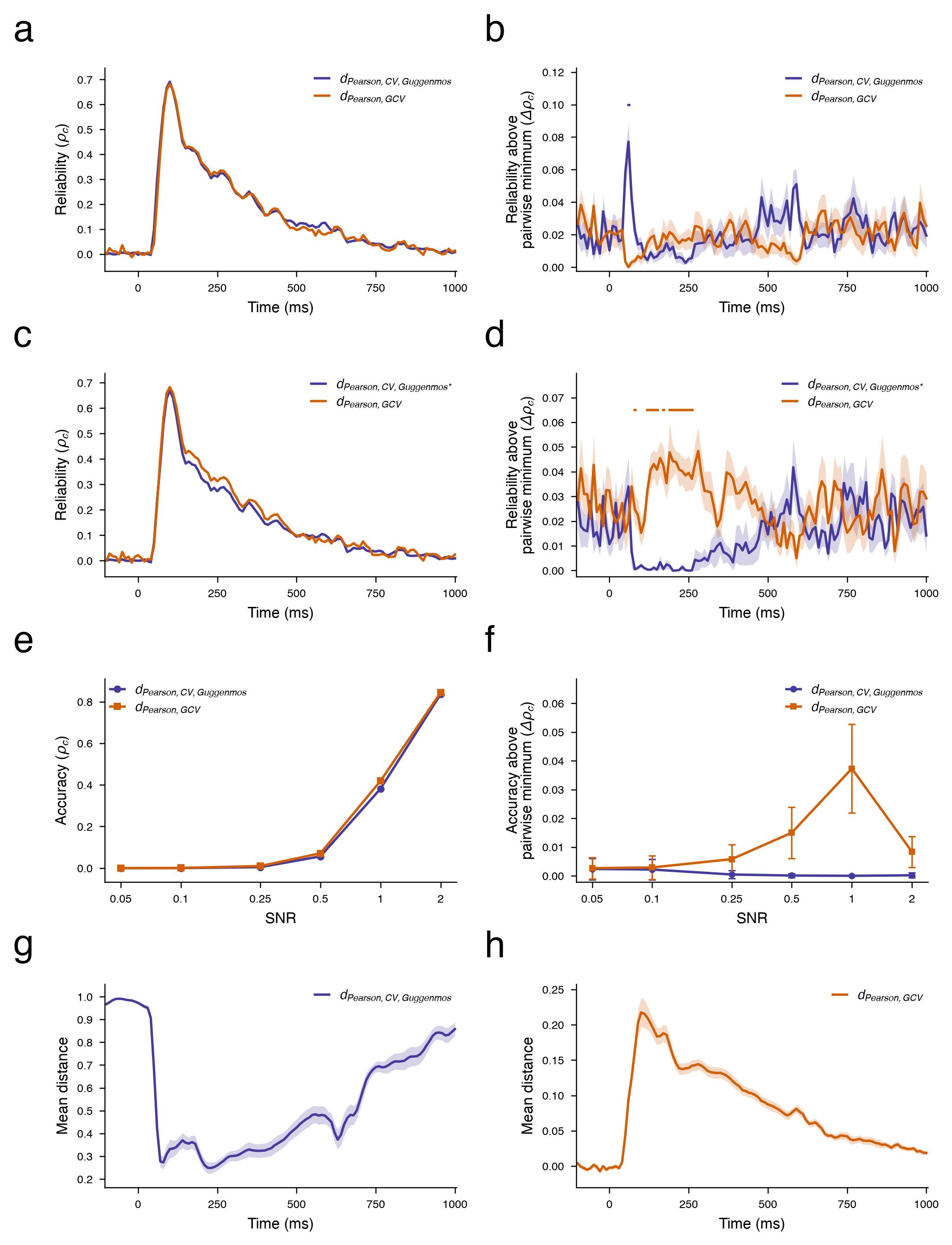}
    \caption{
    Reliability, accuracy and interpretability of cross-validated correlation distances. \textbf{a)} Time-resolved distance reliability (MEG data). \textbf{b)} Between-distances reliability differences (MEG data). For each subject and time point, the minimum between the reliability values of both distances is taken and the difference to that minimum is computed. The group average is shown; the error bars illustrate the standard error of the mean and the top horizontal line illustrates significance ($p < .05$, FWER-corrected, sign permutation test). \textbf{c)} Same as panel \textbf{a}, but with Guggenmos' distance altered by omitting clipping between 0 and 2. \textbf{d)} Same as panel \textbf{b}, but with Guggenmos' distance altered by omitting clipping between 0 and 2. \textbf{e)} Distance accuracy across different SNRs (simulated data). Results for the unaltered Guggenmos' distance are shown; results for the unclipped variation are identical (not shown). \textbf{f)} Accuracy differences between distances (simulated data). For each SNR and simulation iteration, the minimum between the accuracy values of both distances is taken and the difference to that minimum is computed. The group average is shown and the error bars illustrate the standard deviation. \textbf{g)} Mean Guggenmos' distance across time. Error bars illustrate the standard error of the mean. \textbf{h)} Mean G.C.V. correlation distance across time. Error bars illustrate the standard error of the mean. See Appendix C for methodological details.
    }
    \label{fig:Fig2}
\end{figure} 

\section{Cross-validation and within-class correction}

Within-class correction (W.C.C.) is defined as taking the average of all distances computed between trials of different conditions and subtracting the average of all distances computed between trials of the same condition.
\begin{equation}
d_{WCC}(\x, \y) = \frac{1}{N_t^2} \sum_{i=1}^{N_t} \sum_{j=1}^{N_t} d(\x_i, \y_j) 
-\frac{1}{N_t (N_t - 1)} \sum_{i=1}^{N_t} \sum_{j=i+1}^{N_t}[d(\x_i, \x_j) + d(\y_i, \y_j)],
\end{equation}
where \(d\) is a given distance measure and \(N_t\) is the number of trials.

When the number of trials is 2, this is exactly equivalent to generalized cross-validation (G.C.V.):
\begin{equation}
d_{GCV}(\x,\y) = \frac{d(\x_A,\y_B) + d(\x_B,\y_A) + d(\x_A,\y_A) + d(\x_B,\y_B)}{4} - \frac{d(\x_A,\x_B) + d(\y_A,\y_B)}{2}.
\end{equation}
When there are more than 2 trials, the definitions of G.C.V. and W.C.C. diverge: while W.C.C. computes distances between all trials pairwise and averages the distances, G.C.V. averages trials into two partitions before computing distances.

Importantly, both strategies eliminate the bias intrinsic to Euclidean distances in which distances increase with increasing noise (however, note that a random design is also necessary for W.C.C. to be unbiased). With either cross-validation or within-class correction, a true Euclidean distance of 0 will result in a measured distance distributed symmetrically around 0. Within-class correction may even be slightly more accurate at intermediate SNRs, as indicated by our simulations (Figure 3c-d). 

For correlation distance however, bias cannot be completely eliminated without proper modelling \cite{Archer2008,Diedrichsen2018}. Nevertheless, G.C.V. and W.C.C. remain useful tools in that case as well. Indeed, raw correlation distances usually have a timecourse that is hard to interpret and very different from decoding accuracies and other distances, with values initially near 1 and decreasing after stimulus onset, rather than the opposite. As shown by Guggenmos et al. \cite{Guggenmos2018}, this is prevented with W.C.C.: when this correction is applied, the scale is shifted and the overall time course becomes more interpretable. This led to a recommendation from the authors to use this correction, instead of their form of cross-validation, which has the same problem (Figure 2g). We showed that G.C.V. does the same thing (Figure 2h) and therefore that there is no need to rely on within-class correction. 

Generalized cross-validation is also much more accurate than within-class correction. Indeed, as mentioned above, correlation coefficients are intrinsically biased: their value tends towards 0 with increasing noise. Hence, computing many noisy correlations and averaging them is not equivalent to averaging data before computing correlations. In the former case (within-class correction), the estimate's variance is reduced but the bias remains, while in the latter case (generalized cross-validation), both variance and bias are reduced. In our analyses, we observed much greater accuracy for the G.C.V. correlation distance than for the W.C.C. correlation distance (at intermediate and high SNRs; Figure 3g-h), confirming this reasoning. Although we observed increased reliability for the W.C.C. distance, we would argue that accuracy takes precedence over reliability, as ``a confident but incorrect answer is of no use'' (but the reverse is not true). Note that performing W.C.C. after averaging multiple trials into a smaller number of "pseudo-trials" \cite{Guggenmos2018} is less biased than using raw trials and approaches the accuracy of G.C.V., but that G.C.V. reduces the bias maximally.

\begin{figure}[htbp]
    \centering
    \includegraphics[width=0.8\textwidth]{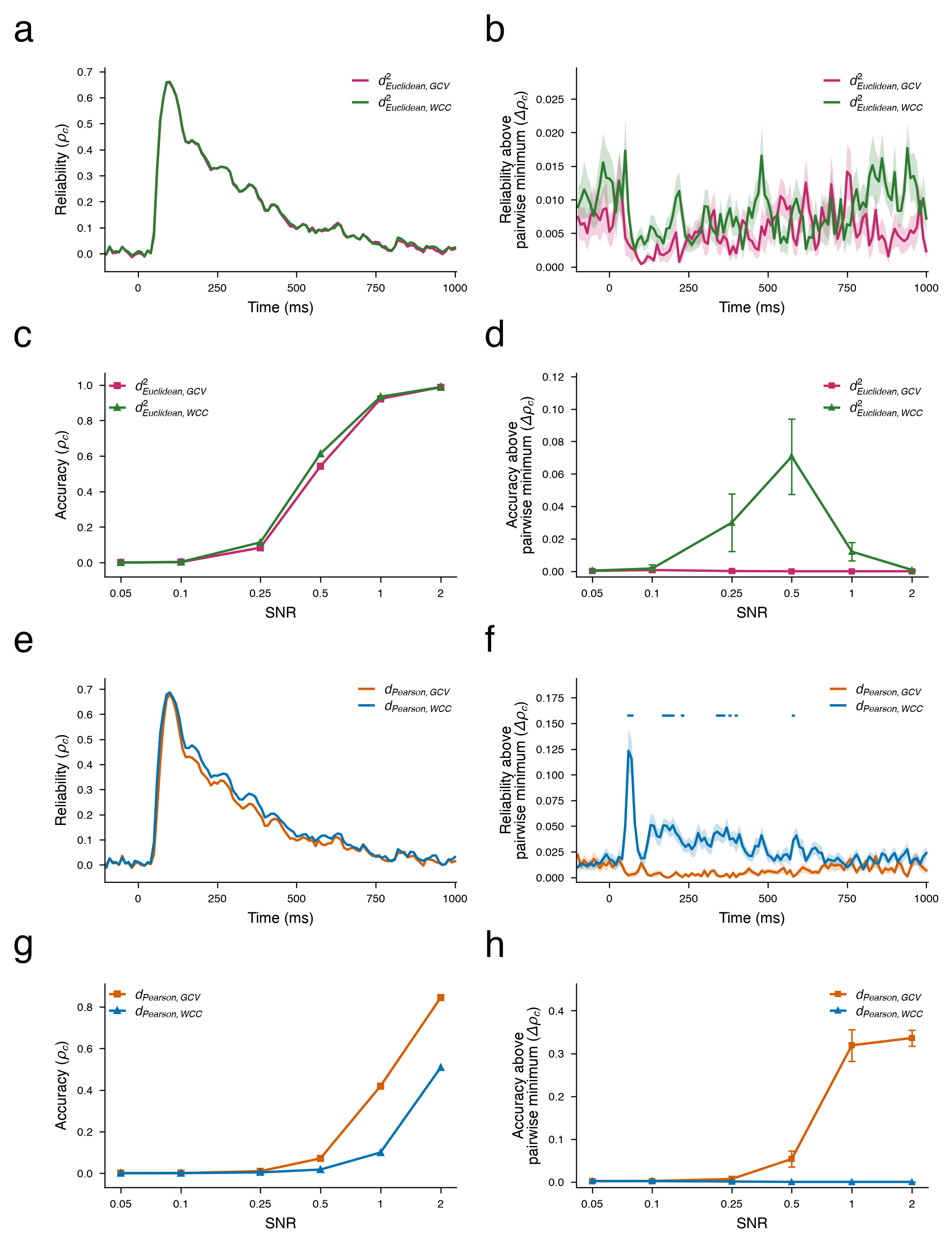}
    \caption{
    Reliability and accuracy of G.C.V. and W.C.C. distances \textbf{a)} Time-resolved reliability of Euclidean distances (MEG data). \textbf{b)} Between-distances reliability differences (MEG data). For each subject and time point, the minimum between the reliability values of both distances is taken and the difference to that minimum is computed. The group average is shown and the error bars illustrate the standard error of the mean. There were no significant differences ($p > .05$, FWER-corrected, sign permutation test). \textbf{c)} Accuracy of of Euclidean distances across different SNRs (simulated data). \textbf{d)} Between-distances reliability differences (simulated data). For each SNR and simulation iteration, the minimum between the accuracy values of both distances is taken and the difference to that minimum is computed. The group average is shown and the error bars illustrate the standard deviation. \textbf{e)} Time-resolved reliability of correlation distances (MEG data). \textbf{f)} Between-distances reliability differences (MEG data). For each subject and time point, the minimum between the reliability values of both distances is taken and the difference to that minimum is computed. The group average is shown; the error bars illustrate the standard error of the mean and the top horizontal line illustrates significance ($p < .05$, FWER-corrected, sign permutation test). \textbf{g)} Accuracy of correlation distances across different SNRs (simulated data). \textbf{h)} Between-distances reliability differences (simulated data). For each SNR and simulation iteration, the minimum between the accuracy values of both distances is taken and the difference to that minimum is computed. The group average is shown and the error bars illustrate the standard deviation. See Appendix C for methodological details.
    }
    \label{fig:Fig3}
\end{figure} 

\section{Conclusions}

In summary, we showed that we can improve the existing formulations of cross-validated (and within-class-corrected) variants of Euclidean and Pearson distances. Our conclusions should also apply to the Mahalanobis and Spearman distances, which are simply the Euclidean distance with prior multivariate noise normalization and the Pearson distance with prior rank transform, respectively. Our formulations led to generally more reliable estimates on MEG data. While it remains to be tested, this increased reliability should also apply to other data modalities. We also hypothesized, backed by statistical reasoning, that our cross-validated correlation distance should be more accurate than both the cross-validated version of Guggenmos et al. \cite{Guggenmos2018} and the within-class-corrected variant, which was confirmed by our simulations. One small caveat to those results is that our generalization of cross-validation cannot be applied if the data partitions are fixed and not random. In such a case, noise may not be orthogonal from one trial to another and the equivalence of between-partition and within-partition distances cannot be assumed. This should not be a problem in the vast majority of studies however, given that conditions are usually presented in a random order. Even in the minority of non-random experimental designs, orthogonal partitions can often be created.

Our work leads to a few recommendations for researchers hoping to use cross-validated distances in their future work. First (notwithstanding the small caveat above), we recommend discontinuing the use of the current form of the cross-validated Euclidean distance and either use its generalized cross-validation variant or the within-class-corrected variant, because they are both more reliable and accurate. If the amount of computation is not a problem, the within-class-corrected Euclidean distance might even be preferable, as our simulations indicate that it's usually more accurate. Furthermore, we recommend discontinuing the use of both Guggenmos et al.'s \cite{Guggenmos2018} implementation of the cross-validated correlation distance and the within-class-corrected correlation distance, in favor of the generalized cross-validated correlation distance, which is more accurate than both. In conclusion, we hope that these updated distance equations lead to increased statistical power and reproducibility of future studies applying distances on noisy data, for RSA or any other purpose.


\begin{appendices}

\newpage
\section{Alternative formulation of the cross-validated Euclidean distance}
\renewcommand{\theequation}{A.\arabic{equation}}
\setcounter{equation}{0}

It can be shown that the cross-validated squared Euclidean distance is equivalent to a sum of between-partition distances. Specifically, the cross-validated squared Euclidean distance is ordinarily defined as:
\begin{equation}
d_{Euclidean,CV}^2(\x,\y) = (\x_A-\y_A)^\top(\x_B-\y_B).
\end{equation}
Expanding the product, we obtain:
\begin{equation}
d_{Euclidean,CV}^2(\x,\y) = \x_A^\top\x_B -\x_A^\top\y_B -\y_A^\top\x_B +\y_A^\top\y_B.
\end{equation}

The squared Euclidean distance between different partitions of the same condition \(x\) is
\begin{equation}
d^2(\x_A,\x_B) = (\x_A-\x_B)^\top(\x_A-\x_B),
\end{equation}
which can be expanded:
\begin{equation}
d^2(\x_A,\x_B) = \x_A^\top\x_A + \x_B^\top\x_B - 2\x_A^\top\x_B.
\end{equation}
Similarly for condition \(y\):
\begin{equation}
d^2(\y_A,\y_B) = \y_A^\top\y_A + \y_B^\top\y_B - 2\y_A^\top\y_B.
\end{equation}
The same can be done for the squared Euclidean distance between different partitions of different conditions:
\begin{equation}
d^2(\x_A,\y_B) = \x_A^\top\x_A + \y_B^\top\y_B - 2\x_A^\top\y_B,
\end{equation}
\begin{equation}
d^2(\x_B,\y_A) = \x_B^\top\x_B + \y_A^\top\y_A - 2\x_B^\top\y_A.
\end{equation}
When subtracting the mean within-condition distance from the mean between-condition distance, many of the expanded terms cancel out, such that we obtain:
\begin{align}
\frac12(d^2(\x_A,\y_B)+d^2(\x_B,\y_A)) -\frac12(d^2(\x_A,\x_B)+d^2(\y_A,\y_B))
&= \x_A^\top\x_B - \x_A^\top\y_B - \y_A^\top\x_B + \y_A^\top\y_B.
\end{align}

The right side of equation A.8 is the same as the right side of equation A.2, so we can confirm that
\begin{equation}
d_{Euclidean,CV}^2(\x,\y) 
= \frac{d^2(\x_A,\y_B)+d^2(\x_B,\y_A)}{2} - \frac{d^2(\x_A,\x_B)+d^2(\y_A,\y_B)}{2}.
\end{equation}

\newpage
\section{Alternative formulation of the cross-validated Pearson distance}
\renewcommand{\theequation}{B.\arabic{equation}}
\setcounter{equation}{0}

We made use of the mathematical relationship between the Pearson correlation and the Euclidean distance when vectors are Z-transformed \cite{Berthold2016} and derived an additional formulation of the cross-validated correlation distance that does not require regularization. We will first recall the definition of a Pearson correlation:
\begin{equation}
r(\x,\y) = \frac{cov(\x, \y)}{\sqrt{var(\x)var(\y)}}.
\end{equation}
Since correlations are not affected by linear transformations of the variables such as z-scoring, and given that the variance of z-scores is 1, we can replace the variables on the right side of the equation with their Z transforms and the terms in the denominator cancel out:
\begin{equation}
r(\x,\y) = cov(Z(\x), Z(\y)),
\end{equation}
where \(Z(x)\) and \(Z(y)\) are centered and scaled versions of the original vectors. The covariance of z-scores can further be reduced to a scaled dot product, where \(n\) is the number of elements in each vector:
\begin{equation}
r(\x,\y) = \frac{Z(\x) \cdot Z(\y)}{n}
\end{equation}

Now, let's look at the squared Euclidean distance, which can be defined as:
\begin{equation}
d^2(\x,\y) = (\x-\y)^\top (\x-\y).
\end{equation}
Following Berthold and Höppner \cite{Berthold2016}, the squared Euclidean distance between Z-transformed vectors reduces to:
\begin{equation}
d^2(Z(\x),Z(\y)) = 2n - 2(Z(\x) \cdot Z(\y)).
\end{equation}
Substituting terms, based on Eq. B.3,
\begin{align}
d^2(Z(\x),Z(\y)) &= 2n(1-r(\x,\y)).
\end{align}
therefore:
\begin{align}
d_{Pearson}(\x,\y) &= \frac{d^2(Z(\x),Z(\y))}{2n}
\end{align}

Thus, based on Eq. B.4, we can define the Pearson correlation distance as such:
\begin{align}
d_{Pearson}(\x,\y) &= \frac{(Z(\x)-Z(\y))^\top (Z(\x)-Z(\y))}{2n}
\end{align}
Finally, this equation lends itself easily to cross-validation:
\begin{align}
d_{Pearson,CV,ours}(\x,\y) &= \frac{(Z(\x_A)-Z(\y_A))^\top (Z(\x_B)-Z(\y_B))}{2n},
\end{align}
where \(A\) and \(B\) are distinct data partitions.

\newpage
\section{Methodological details}
\renewcommand{\theequation}{C.\arabic{equation}}
\setcounter{equation}{0}

\subsection{Simulations}

We adopted a similar approach as the one used by Guggenmos et al. \cite{Guggenmos2018} (see their Figure S5). Specifically, for each SNR level, we ran 500 simulations. Each time, ground truth activation patterns for 306 sensors and 92 conditions were generated by sampling from correlated standard normal distributions. We also added a strong condition-nonspecific pattern (accounting for 5 times the variance of condition-specific patterns). These patterns were replicated across 2 sessions and 20 trials per session. For each trial, zero-mean Gaussian random noise was added, in a proportion to ensure the desired SNR.

\subsection{Reliability and accuracy computation}

Reliability was computed as Lin's Concordance Correlation Coefficient ($\rho_{c}$) between the MEG data from both recording sessions. This coefficient is defined as
\begin{equation}
\rho_c = \frac {2r\,\sigma_x\sigma_y} {\sigma_x^2+\sigma_y^2+(\mu_x-\mu_y)^2},
\end{equation}
where \(r\) is the Pearson correlation coefficient, \(\mu_x\) and \(\mu_y\) are the means of the variables, and \(\sigma_x\) and \(\sigma_y\) are their standard deviations. It measures how well two variables agree by how much the data deviates from a 45 degree line. We used this measure because it is sensitive to both pattern reliability and overall scale shift, as well as location shift \cite{Lin1989}. Furthermore, its values lie between 0 and 1, and it does not require adjusting the scales of variables so that their expected values under the null distribution is always 0 (unlike SSQ reliability), which may require additional assumptions. Reliability was computed for every subject and time point, and was then averaged across subjects. Significant differences between distances were assessed with a sign permutation test across subjects for every time point, and the p-values were adjusted for the Familywise Error Rate across time points (from 50 ms after onset) using a Bonferroni correction \cite{Guggenmos2018}.

Accuracy was computed as Lin's Concordance Correlation Coefficient between measured distance and true distance for each simulation iteration and signal-to-noise ratio. This measure is sensitive to inaccuracies in both the structure of distances (i.e. the representational geometries) and their overall scales.

\end{appendices}

\end{document}